\documentclass{elsart1p}

\usepackage{amssymb,amsmath}
\usepackage{graphicx}
\usepackage[frenchb,english]{babel}
\usepackage{enumerate}
\usepackage{natbib}
\usepackage{lineno}

\usepackage{ifpdf}
\ifpdf
\usepackage[%
  pdftitle={Perturbative treatment of the cosmological N-body problem},%
  pdfauthor={Bruno Marcos},%
  pdfsubject={Vlasovia proceedings},%
  pdfkeywords={N-body simulations, fluid theory, linearized gravity, discreteness effects},%
  pdfstartview=FitH,%
  bookmarks=true,%
  bookmarksopen=true,%
  breaklinks=true,%
  colorlinks=true,%
  linkcolor=blue,anchorcolor=blue,%
  citecolor=blue,filecolor=blue,%
  menucolor=blue,pagecolor=blue,%
  urlcolor=blue]{hyperref}
\else
\usepackage[%
  breaklinks=true,%
  colorlinks=true,%
  linkcolor=blue,anchorcolor=blue,%
  citecolor=blue,filecolor=blue,%
  menucolor=blue,pagecolor=blue,%
  urlcolor=blue]{hyperref}
\fi

\makeatletter
\def\elsartstyle{%
    \def\normalsize{\@setfontsize\normalsize\@xiipt{14.5}}
    \def\small{\@setfontsize\small\@xipt{13.6}}
    \let\footnotesize=\small
    \def\large{\@setfontsize\large\@xivpt{18}}
    \def\Large{\@setfontsize\Large\@xviipt{22}}
    \skip\@mpfootins = 18\p@ \@plus 2\p@
    \normalsize
}
\@ifundefined{square}{}{}
\makeatother

\def\be{\begin{equation}} 
\def\ee{\end{equation}} 
\def\bea{\begin{eqnarray}} 
\def\eea{\end{eqnarray}}  
\def\bean{\begin{eqnarray*}} 
\def\eean{\end{eqnarray*}} 
\def\dd{\partial}

\def\bk{{\bf k}}  
\def\bx{{\bf x}}  
\def\bv{{\bf v}}
\def\br{{\bf r}}

\def\bse{\begin{subequations}}
\def\ese{\end{subequations}}

\def\da{{\dot a}}

\def\lsim{\raise 0.4ex\hbox{$<$}\kern -0.8em\lower 0.62ex\hbox{$\sim$}} 
\def\gsim{\raise 0.4ex\hbox{$>$}\kern -0.7em\lower 0.62ex\hbox{$\sim$}}

\def\bk{\mathbf{k}}

\def\f0N{f_0^{(N)}}
\def\bec{\begin{center}}
\def\eec{\end{center}}

\def\piG{{\pi G}}

\begin{document}

\begin{frontmatter}
\title{Cosmological simulations of structure formation and the Vlasov equation}

\author{Michael Joyce}
\address{Laboratoire de Physique Nucl\'eaire et de Hautes Energies,\\
Universit\'e Pierre et Marie Curie-Paris 6, UMR 7585,\\ Paris, F-75005 France.}

\ead{joyce@lpnphep.in2p3.fr}

\begin{abstract}
In cosmology numerical simulations of structure formation are now of 
central importance, as they are the sole instrument for providing 
detailed predictions of current cosmological models for a whole class 
of important constraining observations. These simulations are essentially 
molecular dynamics simulations of N ($\gg 1$, now up to of order several 
billion) particles interacting through their self-gravity. While their 
aim is to produce the Vlasov limit, which describes the underlying (``cold
dark matter'') models, the degree to which they actually do produce this 
limit is currently understood, at best, only very qualitatively, and there 
is an acknowledged need for ``a theory of discreteness errors''. In this 
talk I will describe, for non-cosmologists, both the simulations and 
the underlying theoretical models, and will then focus on the issue of 
discreteness, describing some 
recent progress in addressing this question quantitatively.
\end{abstract}

\end{frontmatter}

%%%%%%%%%%%%%%%%%%%%%%%%%%%%%%%%%%%%%%%%%%%%
%% MAINMATTER
%%%%%%%%%%%%%%%%%%%%%%%%%%%%%%%%%%%%%%%%%%%%

\section{Introduction}

Cosmology is a branch of physics which has been extremely
dynamic in the last decade or so. This dynamism has come 
essentially from the explosion of new data constraining 
cosmological models, from a whole range of observational techniques. Most
important of all are, without doubt, the observations of the
``three degree Kelvin'' cosmic microwave background radiation
(CMBR): there now exist  detailed maps of the tiny fluctuations 
in the temperature of this radiation as a function of direction 
on the sky,  at an angular resolution well below a degree with a 
sensitivity of less than 100 $\mu$K [\cite{wmap}]. This is
the ``relic'' radiation which decoupled (electromagnetically) from 
matter when the universe was a few hundred thousand years old. 
These maps of temperature fluctuations thus provide very precise 
constraints on the spatial fluctuations in the energy density in 
the universe at this time, which may in turn be related (through 
the Poisson equation, suitably modified in the context of general 
relativity) to the fluctuations in the gravitational potential 
at this time. In the paradigm defined by current cosmological models 
these very low amplitude fluctuations at ``early'' times
develop by gravitational instability, giving rise ultimately to
the highly inhomogeneous universe observed today: the distribution 
of galaxies and galaxy clusters which shows a high degree 
of inhomogeneity with structures extending over a significant fraction 
of the visible universe [see e.g. \cite{book}]. 
This is the problem of ``large scale structure formation'', one
of the central challenges in cosmology today.

The specific model of the universe which cosmologists have
developed to account for current observations posits that 
the dominant fraction (more than eighty per cent) of the 
gravitating matter in the universe is so called ``cold dark matter''
(CDM). This is matter with small velocity dispersion (and therefore 
non-relativistic) which interacts only gravitationally (or almost
only). Further on the scales of interest the approximation of Newtonian 
gravity is very good. Thus the problem of structure formation can 
be reduced, to a good approximation, to that of {\it self-gravitating 
particles in an infinite space evolving from some given initial
conditions (see below)}\footnote{The expansion of the background 
space-time, which is taken into account in cosmological 
simulations, leads only to a very simple modification (see below) 
of the Newtonian equations of motion in a static space-time.}. 
These initial conditions (constrained as we
have noted by the CMBR) are what we can call quasi-uniform,
i.e., they exhibit small deviations from an exactly uniform
distribution of constant density, fluctuations averaged in
a sphere about the mean density typically decaying away as 
some power of the radius.

The CDM particles are, further, usually posited to be microscopic, 
with masses of the order of those of particles in the standard
model of particles physics (i.e. within a few orders of magnitude
of the proton mass). The macroscopic scales on which fluctuations are 
probed by cosmological observations thus contain a huge number of
such particles (e.g. of order $10^{70}$ to $10^{80}$ in standard models).  
This suggests that the relevant dynamics of the CDM can be well 
approximated by the Vlasov limit, taking the particle number $N$ 
to infinity in an appropriate manner. In fact it is assumed canonically 
in cosmology that the correct dynamical description of CDM is given by 
the coupled Vlasov-Poisson equations:
\bse
\label{vlasov-poisson}
\begin{align}
\label{vlasov}
&\frac{\dd f(\br,\bv,t)}{\dd t}+\bv\cdot\nabla f(\br,\bv,t)-\nabla\Phi\cdot \frac{\dd f(\br,\bv,t)}{\dd \bv}=0\\
\label{poisson}
&\nabla^2\Phi(\br,t)=4\piG \left[ \int d^3v  f(\br,\bv,t) - \rho_0 \right]
\end{align}
\ese
where $f(\br,\bv,t)$ is the smooth one-particle phase space density 
(i.e. the mass per unit phase space volume), $G$ the gravitational 
constant, $\Phi$ the gravitational potential and $\rho_0$ the mean
mass density.

In the literature on cosmology and astrophysics the use of
the Vlasov-Poisson system is usually very qualitatively 
justified using simple considerations about time scales
for two body collisions [see e.g. \cite{binney}].
Authors who give derivations 
mostly proceed [e.g. \cite{peebles_80, saslaw_2000}] 
through a truncation of the BBGKY hierarchy for the full N 
particle phase space density, but one by \cite{buchert+dominguez} 
is closely inspired by the approach [see, e.g., \cite{spohn}] probably 
more familiar to the audience of this conference, which uses a coarse-graining 
of the exact single particle ``spiky'' phase space density which
obeys a Liouville equation. These derivations of the Vlasov-Poisson
limit for the purely self-gravitating CDM system in this context 
are not rigorous: they show that the equations
(\ref{vlasov-poisson}) may be obtained in a systematic manner 
when an infinite number of other terms are negligible. 
They do not show rigorously, however, that the corresponding regime 
actually exists, nor that it is the one describing the CDM limit.
We will not address this question here which may, however, be of 
interest to participants at this conference\footnote{See contribution of
M. Kiessling.}.

A central theoretical problem in cosmology is thus the solution
of this set of Vlasov-Poisson equations starting from given 
initial conditions. Analytically they have 
so far proved intractable albeit for some very trivial cases. 
Using various perturbative approaches [see e.g. \cite{peebles_80}], however, 
simple general results can be derived. The most important such result is
that, at leading order in a linearisation of the density and
velocity perturbation fields, the amplitude of the fluctuations 
in the density field, analysed in Fourier space, grow in a simple way, 
{\it independent of the wavelength}, as a function of time\footnote{See 
the contribution of B. Marcos (arXiv:0805.1500) for more detail.}. 
This approach can be extended to higher order, but breaks
down inevitably when the density fluctuations become large.

\section{Dissipationless N body simulations}

Beyond the perturbative regime just described numerical simulations 
are essentially the sole instrument used to model the evolution of CDM, and
hence for extracting detailed predictions of the
distribution of matter in the universe today for confrontation with
data\footnote{We discuss in this paper solely the full three dimensionsal
case as in cosmology. Studies of the analagous one-dimensional system, 
which show qualitatively similar 
features [see, e.g., \cite{rouet-feix, miller-rouet}], may be useful 
for clarifying some of the issues discussed here.}. 
These simulations, however, do not attempt to solve the Vlasov-Poisson 
directly : it is simply not numerically
feasible to do so because of the phase space resolution necessary.
Structures form under gravity by the collapse of overdensities. 
To be useful in cosmology simulations must resolve, at the very least, 
two decades in space and momentum, which translates (in three dimensions)
to treating the long-range interaction on a grid with $10^{12}$ points. 
Even if clever algorithms allow one to reduce the number of operations
at each step to much less than the square of the number of grid points,
this is too much to handle even with current supercomputers\footnote{See 
the contribution of S. Colombi.}.

Instead of tackling the problem directly, cosmologists solve numerically
the much simpler problem of the evolution of N self-gravitating bodies, 
where $N$ is as large a number as feasible. In the first generation of CDM 
simulations in the 1980s $N$ was of the order of a few tens of 
thousands, but it has increased continuously to attain 
almost ten billion in recent simulations! We now describe very briefly
these simulations and outline, qualitatively, some of the results
which they have given. We then turn to the ``problem of discreteness'',
which is simply that of the relation between the results of these
simulations and the solutions of the Vlasov-Poisson system they
attempt to model. 

\subsection{Methods}

We consider what are known as ``dissipationless CDM simulations'', which
are an important class of simulation used to make predictions of the
current CDM cosmological models. They are so called neglect the dissipative 
processes which come into play primarily
because of the gravitational coupling of CDM to ordinary matter. The 
essential features of these N-body simulations (NBS) are the following
[for reviews see e.g. \cite{bagla, bertschinger}] :

\begin{itemize}

\item  One considers $N$ particles in a cubic box of side $L$
with {\it periodic boundary conditions}. 

\item The equations of motion which are solved numerically are: 
\be
\label{full-ev}
\ddot \bx_i+2\frac{\da}{a}\dot \bx_i=-\frac{1}{a^3}\sum_{i\ne j} 
\nabla_{x_i} V(|\bx_i-\bx_j|),
%\frac{G m (\bx_i-\bx_j)}{|\bx_i-\bx_j|^3},
\ee
where $\bx_i$ is the ``comoving'' position of the $i$-th particle, 
and $a(t)$ is the ``scale factor'' for the cosmological model 
considered. The latter is a monotonically increasing function
of the time, and grows by a factor of $50-100$ in a typical 
simulation. The two-body potential $V(|\bx_i-\bx_j|)$ 
of the particles is the Newtonian potential, with a 
regularisation of the singular ultra-violet behaviour
at small scales. By ``small'' scales we
mean that the characteristic scale of the smoothing $\varepsilon$ 
is much smaller than the initial mean interparticle 
distance\footnote{The exact form of the smoothing used depends on the code,
and the canonical choice has evolved. Current codes typically have the
exact Newtonian potential above $\varepsilon$. 
%The reason for the use of this smoothing is often roughly motivated 
%physically.
}. 

\item The initial positions and velocities of the particles
  are determined from the initial conditions (IC) of the theoretical
model using the so-called ``Zeldovich approximation''. This 
is an approximation to the evolution of a self-gravitating fluid which 
gives the displacements of fluid elements {\it from a uniform
state}, and their velocities, in terms of the local 
gravitational field. The latter may be determined, and thus 
the displacement and velocity fields, for any set of 
initial density fluctuations (using the Poisson equation).
The theoretical density fluctuations are canonically Gaussian, 
and thus fully specified by their power spectrum\footnote{This
is equivalent, up to a normalization, to what is 
called the ''structure factor'' in solid state physics.}(PS). 
A realization of this PS is thus used to generate the 
displacements and velocities, and these are applied to
the particles of the N body simulation, placed in a
``uniform'' configuration usually taken to be a simple 
cubic lattice (or sometimes ``glassy'' configuration).
 
\end{itemize}
 
\subsection{Results}
Different approaches to characterising the results of N body 
simulations may be found in the cosmological literature. 
While the early literature concentrated almost exclusively
on two point correlation properties measured in real space,  
subsequently reciprocal space properties (notably the PS
which is simply the Fourier transform of the two point correlation 
function) became more widely used. These extensive (and continuing)
studies have led essentially to various phenomenological ansatzes 
which express the final two point correlation properties 
{\it as a functional of the initial two point correlation properties}
[\cite{peacock+dodds}]. 
More recently an alternative description of results in terms of a 
different phenomenological description, the so-called
``halo-model'' has become popular [for a review see
\cite{cooray+sheth}]. This is based on the 
fact that the structures formed in simulations are observed to be 
well described as a collection of virialised clumps, whose masses 
and spatial correlations may be determined from the 
initial conditions, but  whose internal organisation seems 
to have universal properties (i.e. independent of the initial 
conditions).

\section{The problem of discreteness}

As in any numerical simulation errors arise from the 
discretisation of the equations of motion intrinsic
to the method. This is {\it not} what we mean by
the problem of discreteness in cosmological NBS,
which arises from the {\it intrinsic} difference between the
simulated N-body system and the physical system of
a much larger number of particles, which is supposed
to be described, as we have discussed above, by the 
Vlasov-Poisson equations. Concretely the question
can be put as follows: {\it what is the difference, at any given 
time and spatial scale, between, say, a two-point 
correlation function measured in a perfect N body
simulation and the same quantity in the modelled
Vlasov-Poisson system?} This question is, perhaps surprisingly,
one which has been relatively neglected by cosmologists,
and several of the few controlled studies which exist 
[see e.g. \cite{splinteretal}] raise questions about the 
quite optimistic assumptions widely made by simulators 
in linking their results of theory. It has become an issue 
of increasing importance in the last few years --- and will take on
more importance inevitably in the coming years --- as
much greater precision predictions from simulations is 
now required to exploit observational advances. 

In work over the last two years we have tried to approach
this question in a systematic manner. Our work can be 
divided into three parts, the first two of which are 
essentially completed while the third remains under investigation.

\subsection{Discreteness in initial conditions}
It is evidently important to understand the relation between
the continuous theoretical model and the discrete simulated 
system already at the starting time. We have thus 
considered the question of the relation between the 
statistical properties of the density fluctuations
in the initial conditions(IC) and those of the theoretical
input model. As explained above, the IC
are generated by displacing the particles off a perfect
lattice in a manner prescribed by an input PS. 
Using a general formalism developed in \cite{andrea-pre}, we have determined
precisely [\cite{mj-bm-IC}] the PS of the IC of
cosmological simulations. To a first approximation
this is essentially just the sum of the PS of the 
underlying lattice, the input PS and a convolution term.
For wavenumbers below that corresponding to the Nyquist frequency 
of the lattice, the full PS is well approximated by the input one.
In direct (real) space the situation is very different
because the Fourier transform of the lattice PS is
a highly delocalised function, which always dominates
over that of the input PS for sufficiently low amplitude
of the latter. This means that, in the limit of low
amplitude, the discretised system becomes a poor representation
in real space of the continuous model, while remaining
good in most of reciprocal space. Cosmologists usually
analyse the initial conditions only in reciprocal space,
and {\it a priori} it is not clear whether our finding
has dynamical relevance. It turns out, as we will now
explain, that it does.

\subsection{Early time evolution : A perturbative approach }

NBS begin, by construction, in the regime in which the
Zeldovich approximation is valid, which means that the particles 
are initially close to the lattice configuration. In 
this case we have found\footnote{See the contribution of 
B. Marcos (arXiv:0805.1500) for 
a fuller account.} that one can generalise usefully the standard procedure
employed to describe the perturbations to a crystal in solid
state physics: one linearizes the force in the displacements
and determines the eigenmodes and eigenvalues of the $3N \times
3N$ matrix. Each such eigenmode is labelled by a reciprocal
lattice vector $\bk$, and the sign of the eigenvalues determines
whether a mode grow or oscillates. We have shown 
[\cite{PRL, marcos_06}] how one can recover the equivalent
perturbative regime of the fluid limit derived from the 
Vlasov-Poisson system from such a treatment, in the limit
$k\ell \rightarrow 0$ (where $k$ is the modulus of $\bk$,
and $\ell$ is the lattice spacing). Having recovered this
limit we can then quantify precisely
the differences between the evolution of the continuous model
and of its N body discretisation, i.e., the 
discreteness effects in the regime of validity of the 
perturbative approximation. One important result which 
is found is that such differences grow in time, i.e., the
N body dynamics always diverge from the Vlasov dynamics
for sufficiently long times. This is probably a result 
unsurpising to many of those at this conference, but it
was unknown in the context of simulations of gravity
in cosmology: the temporal duration of a simulation
is not a parameter which has been considered in testing
for discreteness effects\footnote{Indeed when such tests are
performed by studying numerically the N dependence of
results [see e.g. \cite{diemandetal}], the temporal length of the 
simulation is actually changed in a manner fixed by 
a criterion built into the standard numerical package 
now used for generating initial conditions.}. We note that
in the limit of low amplitude of the initial PS, this 
perturbative approximation for the dynamics remains
valid for an arbitrarily long in time. Thus the divergence
in this former limit of the discretised IC from the 
theoretical IC, which we described above, translates 
also into a divergence of the evolved discrete system
from its continuum counterpart.

%Another result is that the differences depend on scale,
%with smaller scales diverging more rapidly from the
%Vlasov behaviour.
\subsection{Evolution in the non-linear regime}

All cosmological NBS aim to go well beyond the perturbative regime,
and ultimately it is the role of discreteness in this regime
which we must understand. The problem is evidently intractable
analytically insofar as it poses an even more difficult one that
the (analytically intractable) Vlasov-Poisson system. One can, however, 
approach the problem numerically in various ways. Controlled tests for 
effects explicitly due to discreteness in the system may be defined.
For example, one can investigate the role of interactions between
nearest neighbour particles which should become subdominant in the
Vlasov limit. It turns out [\cite{Baertschigeretal_apj}]
that in NBS, in 
the phase immediately after the perturbative one described above, the 
evolution of the first non-linear correlations (which develop at scales 
well below $\ell$) is very well described by considering {\it only} two body 
interactions, and thus clearly the system {\it at these (small) scales} 
is not described by the Vlasov limit. 
%At later times, however, it is
%believed widely by cosmological simulators that results at these same
%scales 
In his contribution B. Marcos briefly outlines some work in progress,
aimed at addressing the quantification of discreteness effects at 
much later times: the idea is to {\it bound below} discreteness effects
in a given simulation by looking at the dispersion in measured
quantities (e.g. the PS) for discretisations {\it at fixed N} which differ only
in the initial particle distribution used to represent uniformity.
From the study of the evolution of such dispersion we should be able
to get insight into how discreteness effects become more or less
important as strong non-linearity sets in.

\section{Conclusions}

The question of the relation between the dynamics
of a finite number of self-gravitating particles and the Vlasov
limit of the same system is an open problem, which is of
both theoretical and practical importance in cosmology.
While we have managed to essentially solve this problem 
in a specific regime --- early times/low amplitude fluctuations ---
it remains a great challenge to find useful approaches to
the more general problem. We have briefly described some
possible such approaches and some results, which open 
avenues for further investigation. It would certainly be
interesting to see if analagous problems encountered in
other contexts as described at this conference may 
suggest other approaches or give insight into the problem.

%%%%%%%%%%%%%%%%%%%%%%%%%%%%%%%%%%%%%%%%%%%%%%%%
%% BACKMATTER
%%%%%%%%%%%%%%%%%%%%%%%%%%%%%%%%%%%%%%%%%%%%%%%%

%
%\begin{theacknowledgments}
I thank T. Baertschiger, A. Gabrielli, B. Marcos and
F. Sylos Labini for collaboration on the work reported here. 
%I also thank them to
%  have read and for comments on this manuscript.
%\end{theacknowledgments}

%\bibliographystyle{aipproc}   % if natbib is available
%\bibliographystyle{aipprocl} % if natbib is missing

%%%%%%%%%%%%%%%%%%%%%%%%%%%%%%%%%%%%%%%%%%%
%% You probably want to use your own bibtex database here
%%%%%%%%%%%%%%%%%%%%%%%%%%%%%%%%%%%%%%%%%%%
%\bibliography{thesis2}

\begin{thebibliography}{15}
\expandafter\ifx\csname natexlab\endcsname\relax\def\natexlab#1{#1}\fi
\providecommand{\enquote}[1]{``#1''}
\expandafter\ifx\csname url\endcsname\relax
  \def\url#1{\texttt{#1}}\fi
\expandafter\ifx\csname urlprefix\endcsname\relax\def\urlprefix{URL }\fi
\providecommand{\eprint}[2][]{\url{#2}}

\bibitem[Bagla \& Padmanabhan(1997)]{bagla}
J.~S. Bagla, and T.~Padmanabhan, \emph{Pramana} \textbf{49}, 161--192 (1997),
  \eprint{astro-ph/0411730}.

\bibitem[Baertschiger et~al.(2004)]{Baertschigeretal_apj}
T.~Baertschiger, M.~Joyce and F.~Sylos~Labini, 
\emph{Astrophys.J. Lett.} \textbf{581}, L63-L66, 2002.

\bibitem[Bertschinger(1998)]{bertschinger}
E.~{Bertschinger}, \emph{Annu. Rev. Astron. Soc.} \textbf{36}, 599--654 (1998).

\bibitem[Binney \& Tremaine(1987)]{binney}
J.~Binney, and S.~Tremaine, \emph{Galactic Dynamics}, Princeton Series in
  Astrophysics, Princeton, 1987.

\bibitem[Buchert \& Dominguez(2004)]{buchert+dominguez}
T.~Buchert and A.~Dominguez, \emph{Astron.Astrophys.} \textbf{438} ,443---460 (2005) 

\bibitem[Cooray \& Sheth(2002)]{cooray+sheth}
A.~Cooray and R. Sheth, \emph{Phys.Rept.} \textbf{372}, 1-129 (2002), 
\eprint{astro-ph/0206508}.

\bibitem[Diemand et~al.(2004)]{diemandetal}
J.~Diemand, B.~Moore, J.~Stadel, and S.~Kazantzidis, \emph{Mon. Not. Roy.
  Astron. Soc.} \textbf{348}, 977 (2004), \eprint{astro-ph/0304549}.

\bibitem[Gabrielli(2004)]{andrea-pre}
A.~Gabrielli, \emph{Phys. Rev.} \textbf{E70}, 066131 (2004), 

\bibitem[Gabrielli et~al.(2005)]{book}
A.~Gabrielli, F.~Sylos~Labini, M.~Joyce, and L.~Pietronero, \emph{Statistical
  Physics for Cosmic Structures}, Springer-Verlag, Berlin, 2005.

\bibitem[Joyce \& Marcos(2004)]{mj-bm-IC}
M.~Joyce and B.~Marcos, \eprint{astro-ph/0410451}.

\bibitem[Joyce et~al.(2005)]{PRL}
M.~Joyce, B.~Marcos, A.~Gabrielli, T.~Baertschiger, and F.~Sylos~Labini,
  \emph{Phys. Rev. Lett.} \textbf{95}, 011304 (2005).

\bibitem[Marcos et~al.(2006)]{marcos_06}
B.~Marcos, T.~Baertschiger, M.~Joyce, A.~Gabrielli, and F.~Sylos~Labini
  (2006), \emph{Phys. Rev.} \textbf{DD73} 103507 (2006). 

\bibitem[Miller \& Rouet(2002)]{miller-rouet}
B. N.~Miller and J.-L.~Rouet, \emph{Phys. Rev.}
\textbf{E65}, 056121 (2002).

\bibitem[Peacock \& Dodds(1996)]{peacock+dodds}
J.~Peacock and S. Dodds, \emph{Mon. Not. Roy.
  Astron. Soc.} \textbf{280}, L19 (1996), \eprint{astro-ph/0303031}.

\bibitem[Peebles(1980)]{peebles_80}
P.~J.~E. Peebles, \emph{The Large-Scale structure of the Universe}, Princeton
  University Press, Princeton, 1980.

\bibitem[Rouet et al.(1990)]{rouet-feix}
J.-L.~Rouet, M. R.~Feix and M.~Navet, \emph{Vistas in Astronomy}, 
\textbf{33}, 357--370.

\bibitem[Saslaw(2000)]{saslaw_2000}
W.~Saslaw, \emph{The Distribution of the Galaxies}, Cambridge University Press,
Cambridge, 2000.

\bibitem[Splinter et al.(1998)]{splinteretal}
R.~Splinter, A.~Melott, S. Shandarin and Y. Suto \emph{Astrophys.J.} \textbf{497}, 38---61 (1998)

\bibitem[Spohn(1991)]{spohn}
H.~Spohn, \emph{Large Scale Dynamics of Interacting Particle},
Springer Verlag (Berlin), 1991.

\bibitem[WMAP collaboration(2006)]{wmap}
G. Hinshaw et al.(the WMAP collaboration), astro-ph/0603451 

\end{thebibliography}

\end{document}